\documentclass[conference]{IEEEtran}
\IEEEoverridecommandlockouts
\usepackage{cite}
\usepackage{amsmath,amssymb,amsfonts}
\usepackage{bm}
\usepackage{algorithm}
\usepackage{algorithmic}
\usepackage{graphicx}
\usepackage{textcomp}
\usepackage{xcolor}
\usepackage{verbatim}

\def\BibTeX{{\rm B\kern-.05em{\sc i\kern-.025em b}\kern-.08em
    T\kern-.1667em\lower.7ex\hbox{E}\kern-.125emX}}

\newtheorem{lemma}{Lemma}

\begin{document}
\title{Parallel Delay-Doppler Estimation via Order-Reversed Two-Stage Prony Method\thanks{This work was supported in part by JSPS KAKENHI Number JP23K26104 and JP23H00474.}
}

\author{
\IEEEauthorblockN{Yutaka Jitsumatsu}
\IEEEauthorblockA{\textit{Deptartment of Informatics,} \\
\textit{Kyushu University, Japan}\\
jitsumatsu@inf.kyushu-u.ac.jp}
\and
\IEEEauthorblockN{Liangchen Sun}
\IEEEauthorblockA{\textit{Department of Information Science and Technology} \\
\textit{Kyushu University, Japan}\\
sun@me.inf.kyushu-u.ac.jp}
}

\maketitle

\begin{abstract}

This paper proposes a Prony-based parallel two-stage method for delay–Doppler estimation in OTFS systems. 
By performing delay-first and Doppler-first estimations in parallel and fusing the results, the method resolves ambiguities caused by similar path characteristics. 
The simulation results demonstrate the superior accuracy and robustness of the proposed method under various conditions.
This method provides a promising solution for future applications such as Vehicle-to-Vehicle (V2V) and Integrated Sensing and Communication (ISAC).

\end{abstract}


\section{Introduction}

Doubly selective fading in high-mobility channels poses serious challenges to both communication and sensing~\cite{Delay-Doppler-Communications}. 
Signals often include multiple Doppler-shifted components, making symbol detection difficult. 
Orthogonal Frequency-Division Multiplexing (OFDM) can mitigate Doppler effects with methods such as the basis expansion model (BEM)~\cite{Hlawatsch-Book}, but it is computationally expensive. 
Recently, Orthogonal Time Frequency Space (OTFS) modulation~\cite{OTFS, OTFS-BITS, OTFS-BITS2} has gained attention as a low-complexity~\cite{Raviteja2018} and robust solution for doubly selective channels~\cite{Lin_ODDM}. 
However, accurate delay-Doppler estimation remains challenging when the parameters do not align with the integer grid, due to energy leakage and the resulting complex computations~\cite{ zegrar2024otfs,zhang2023radar,zacharia2023fractional,li2022novel, muppaneni2023channel,li2024grid,wang20222d,sheng2023time}.

We have recently proposed a novel delay–Doppler estimation method based on a two-stage Prony approach~\cite{YutakaPIMRC2025}. This method first estimates the Doppler shifts and subsequently estimates the delays associated with each Doppler, providing accurate estimation in low-noise environments.
By exploiting the symmetry between the time and frequency domain representations, a frequency-dual counterpart can be formulated. In this counterpart, the Fourier transform of the received signal is used as the input, and the delays are estimated first, followed by the Doppler shifts for each delay.
In this paper, we employ both the Doppler-first and delay-first methods and combine their delay–Doppler estimates, thereby achieving improved estimation accuracy.

\section{Problem Formulation}
This section presents the continuous-time OTFS signal with a pilot, the doubly selective channel, and the corresponding discrete-time received samples. 
The OTFS delay-Doppler plane is first defined on an $(N,M)$ grid, which is then extended to $(N+2,M+2)$ by adding two samples in both the delay and Doppler dimensions.

\subsection{Transmitted Signal}
We consider an OTFS system with $M$ subcarriers and $N$ time slots per frame.
Let $T$ denote the slot duration, yielding a subcarrier spacing of $1/T$, a frame duration of $NT$, and a bandwidth of $M/T$.
The delay-Doppler (DD)-domain signal is $X_\mathrm{DD}[k,\ell]$ with $k \in [0,N-1]$ and $\ell \in [0,M-1]$, assumed periodic in both indices.
The corresponding time-domain symbol is obtained via the inverse discrete Zak transform (IDZT)~\cite{OTFS-BITS2} as 
$x_\mathrm{TD}[n M + \ell ] =
\frac1N \sum_{k=0}^{N-1} X_\mathrm{DD}[k,\ell] e^{-j\frac{2\pi}N nk}
$. The transmitted signal is then given by
\begin{align}
s(t) = \sum_{n=0}^{N-1} \sum_{\ell=0}^{M-1}
x_\mathrm{TD}[n M + \ell ] p_\mathrm{T} (t- (nM+\ell) \frac{T}{M} ),
\label{s(t)}
\end{align}
where $p_\mathrm{T}(t)$ denotes the transmitter's pulse shape.
We consider channel estimation using a pilot signal defined by $X_\mathrm{DD}[0,0]=1$ and $X_\mathrm{DD}[k,\ell]=0$ for all other $(k,\ell)$.

Compared to conventional rectangular pulses~\cite{ raviteja2019orthogonal,zhang2021modulation,keskin2021radar,gaudio2020effectiveness,shi2023integrated, ranasinghe2024fast}, a key technique in this paper is the careful design of the pulse shape. 
Typically, a sinc pulse or a root-Nyquist pulse (such as the Root Raised Cosine~\cite{sun2025design}) can be used as $p_{\rm T}(t)$. 
In this paper, we adopt
\begin{align}
p_\mathrm{T}(t) = \frac{\sin (\pi M t/T)}{\pi t} \cdot e^{-j \pi t/T},
\end{align}
which is a frequency-shifted ideal sinc pulse resulting from the DFT indexing $-M/2,\ldots,M/2-1$ (with even $M$), centering the spectrum at $-1/(2T)$ instead of $0$.
Extension to root-Nyquist pulses is left for future work.
On this basis, we further effectively approximate the above transmit signal (1) as a Dirac waveform, defined for $t \in [0, (N-1)T]$,
\begin{align}
\frac{1}{T} \sum_{m=-M/2}^{M/2-1} e^{j m \frac{2 \pi}{T} t}
= \frac{\sin ( M \pi t/T)}{T \sin (\pi t/T)} \cdot e^{-j \pi t/T}.
\label{dirichlet_waveform}
\end{align}

\begin{figure}
    \centering
    \includegraphics[width=1\linewidth]{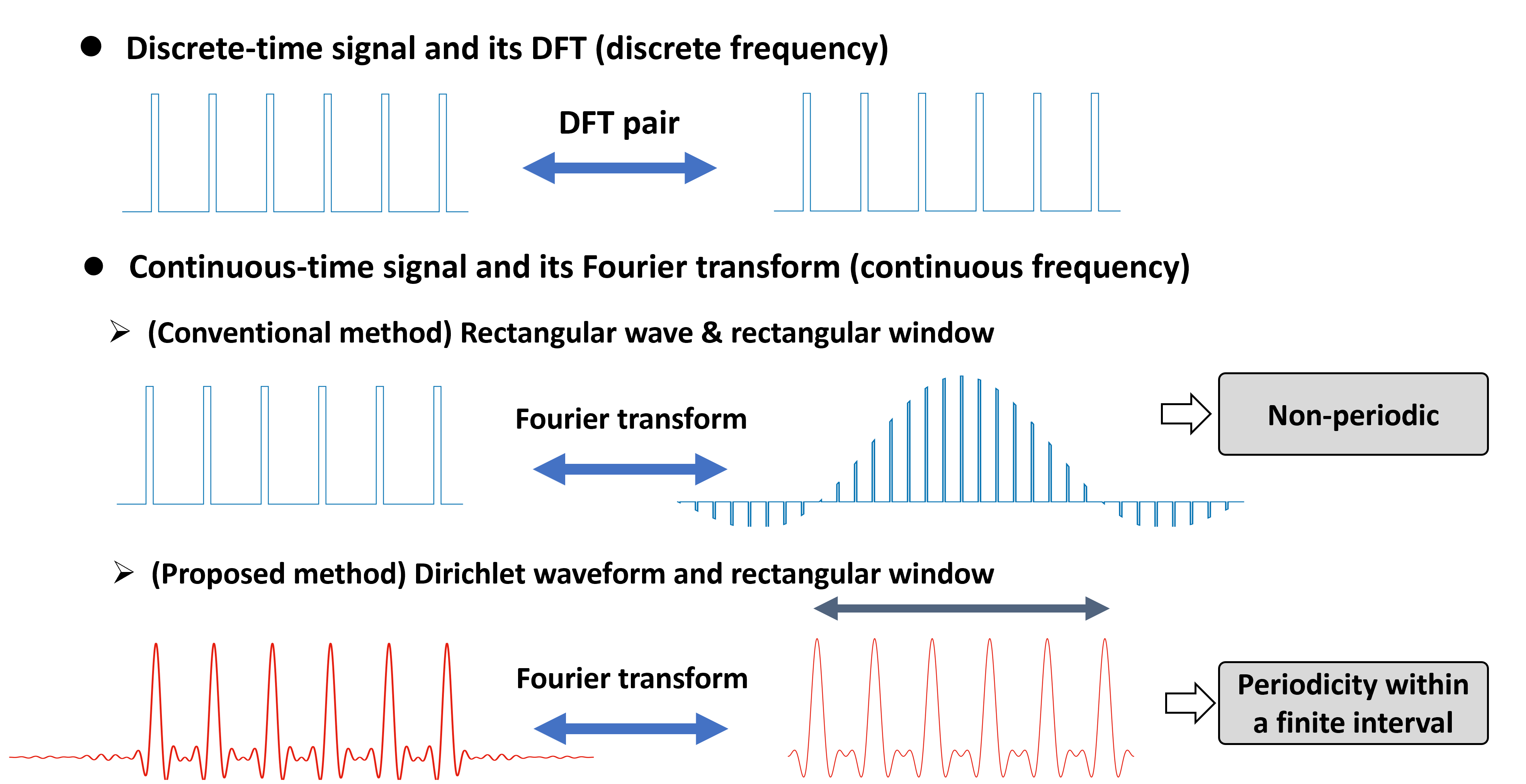}
    \caption{Comparison of rectangular and Dirichlet waveform approximations to a discrete-time signal in time and frequency domains.}
    \label{fig:dirichlet_vs_sinc_pulses}
\end{figure}

Fig.~\ref{fig:dirichlet_vs_sinc_pulses} illustrates approximations of a discrete-time transmitted signal using rectangular pulses and the Dirichlet waveform.
The rectangular pulses align well with the discrete-time impulses in the time domain, but they lose periodicity in the frequency domain.
In contrast, the Dirichlet waveform exhibits oscillations around the impulses due to its sinc-like shape, yet it preserves periodicity within a finite interval.
This latter property is fully exploited by the proposed two-stage Prony method, which is why we employ the Dirichlet waveform as the pulse shape.


\subsection{Receive signal}\label{sect:channel_model}
Consider a multi-path channel with Doppler shifts, where $P$ is the (unknown) number of propagation paths, typically much smaller than $NM$, the signal-space dimension.
Let $t_{d,\max}$ and $f_{D,\max}$ denote the maximum delay and maximum Doppler shift, respectively.  
For the $p$-th path, let $\alpha_p$, $t_{d,p}$, and $f_{D,p}$ denote the attenuation factor, time delay, and Doppler shift.  
We assume that $\alpha_p$ follows a complex Gaussian distribution and $t_{d,p}$ and $f_{D,p}$ satisfy $0 < t_{d,p} < T, -\frac{1}{2T} < f_{D,p} < \frac{1}{2T}$. 
Under this condition, the received signal is defined as
\begin{align}
    r(t) = \sum_{p=1}^P \alpha_p s(t - t_{d,p}) e^{j 2\pi f_{D,p} t} + z(t),
    \label{r(t)}
\end{align}
where $z(t)$ denotes additive white Gaussian noise (AWGN).  
Due to delay and Doppler shifts, the effective periodicity interval is shortened by up to $2T$ and $2/T$, respectively.

\begin{figure}
    \centering
    \includegraphics[width=1\linewidth]{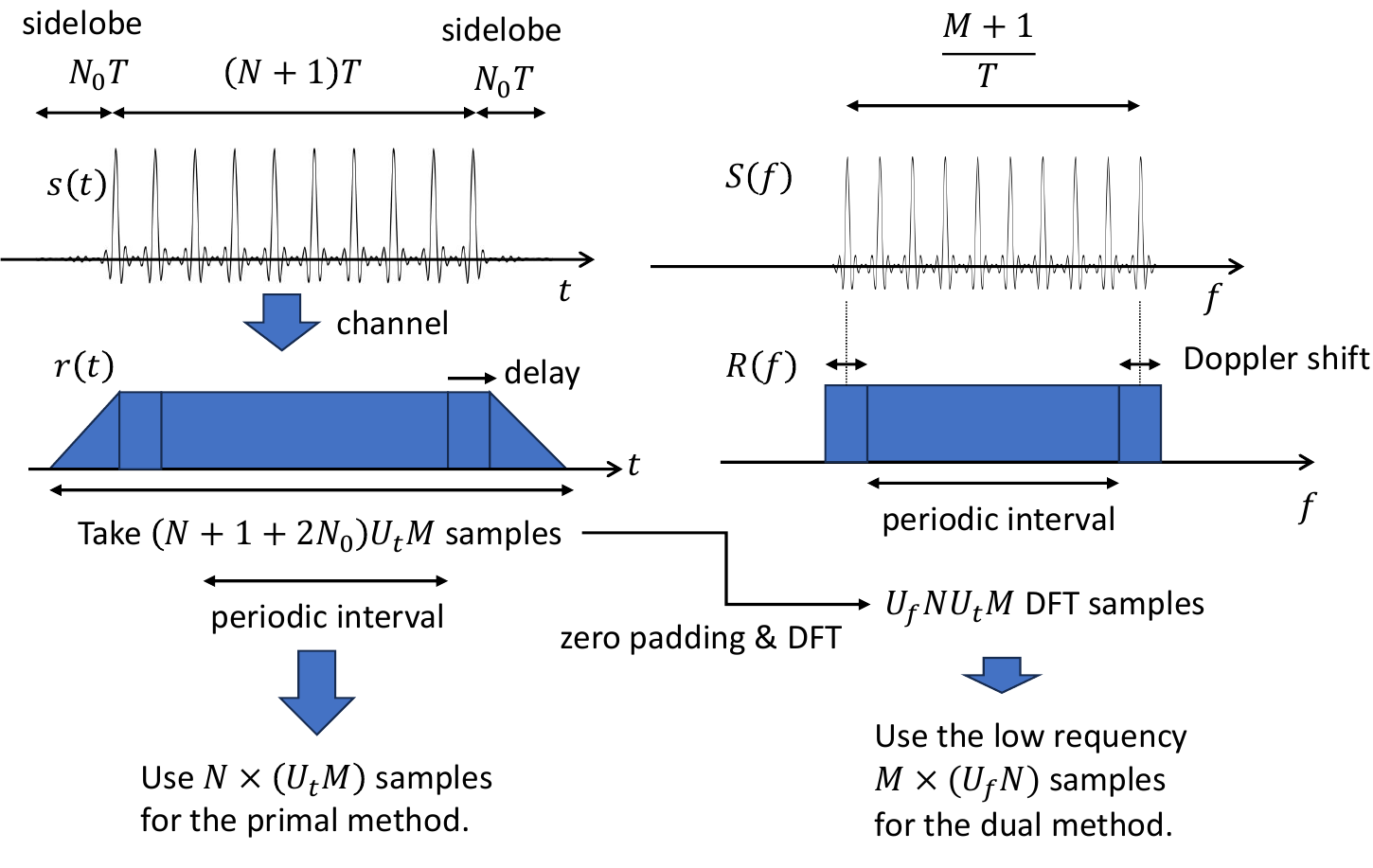}
    \caption{Time-domain and frequency-domain samples of OTFS signaling with a pilot symbol}
    \label{fig:Received_Signal}
\end{figure}

Two additional samples, as Fig.~\ref{fig:Received_Signal} shows, are introduced both in the time and frequency domains, yielding an effective size of $(N+2, M+2)$. 
The received signal is then sampled with interval $T_s = T/(U_t M)$, where $U_t$ is the upsampling factor, sufficiently set to 2, yielding
\begin{align}
    r_\mathrm{TD}[\ell] 
    = \int r(t) p_\mathrm{R}(t-\ell T_s) dt \approx r(\ell T_s).
\end{align}
Here, $p_\mathrm{R}(t)$ is a low-pass filter before sampling at a rate of $T_s$. 
This approximation holds because $f_{D,\max}$ has been assumed smaller than $1/T$, which is smaller than $1/T_s$ by a factor of $U_t M \gg 1$.

Then, let $R(f)$ denote the Fourier transform of $r(t)$, and define its samples as
$R_{\mathrm{FD}}[k] = R\left( k \varDelta f \right)$, where $\varDelta f =\frac1{U_f NT}$ is the frequency bin,
and $U_f$ denotes the upsampling factor in the frequency domain. Throughout this paper, we assume $U_f = 2$.
The frequency-domain samples are thus computed as 
\begin{align}
    R_{\mathrm{FD}}[k] 
    &=
    \int r(t) e^{j 2\pi \frac{k}{U_f NT} t } dt \notag \\
    &\approx
    T_s\sum_{\ell} r(\ell T_s) e^{j 2\pi \frac{k}{U_f NT} \ell T_s } \notag\\
    &=
    T_s\sum_{\ell} r_\mathrm{TD}[\ell] e^{j 2\pi \frac{k\ell}{U_t U_f N M}  } .
\end{align}
The approximation is accurate due to time-domain upsampling. 
The summation range of $\ell$ captures the significant portions of the sum-of-sinc pulse, as shown in Fig.~\ref{fig:dirichlet_vs_sinc_pulses}. 
Let $N_0 T$ denote the tail portions of $s(t)$, with $N_0 = 2$, which is sufficient in our case. 
Adding two extra peaks extends the total support to $[-N_0 T,\, (N+1+N_0)T]$, yielding $(N+1+2N_0) U_t M$ samples. 
DFT coefficients are obtained by zero-padding this sequence to length $U_f U_t N M$.

\section{Proposed Methods}

In this chapter, we provide a detailed introduction to the proposed parallel delay-Doppler estimation approach, which includes two submethods. The first is the Doppler-first method (Section~\ref{sect:primal}), which extends our earlier work~\cite{YutakaPIMRC2025} by introducing an oversampling factor $U_t>1$. In Prony's method, the number of paths, denoted by $\hat P$, must be specified in advance. In~\cite{YutakaPIMRC2025}, we applied Prony's method with $\hat P=1,2,\ldots, N-1$ and selected the best value using information criteria. In this paper, however, we fix $\hat P=N-1$ for the Prony method. 
The second is the symmetric delay-first method (Section~\ref{sect:dual}), in which we set $\hat P=M-1$. The final results, including the estimation of the number of paths, are obtained by combining the two methods.

Since the finite periodicity of $r_{\rm TD}[\ell]$ is broken in the tails due to delays, we discard the tail samples in this section and relabel the remaining samples $t \in [T, (N+1)T]$ as $0,1,\ldots,NU_tM-1$. This preserves periodicity without affecting the estimates of the delay $t_{d,p}$ and Doppler shift $f_{D,p}$.

\subsection{Primal Two-Stage Prony Method: Doppler-First}\label{sect:primal}

\begin{lemma}
Let $\bm{R} = (R_{n,\ell})$ be an $N \times (U_t M)$ matrix with
\begin{align}
R_{n,\ell} = r_{\rm TD}[nU_tM+\ell],
\end{align}
and let $\bm{E} = (E_{n,p})$ and $\bm{V} = (V_{p,\ell})$ be
\begin{align}
E_{n, p} &= e^{j2\pi f_{D,p} nT}, \\
V_{p,\ell} &= \alpha_p s\left( \ell T_s - t_{d,p} \right) e^{j2\pi f_{D,p} \ell T_s}.
\end{align}
Then, the received signal can be written as the matrix product
\begin{align}
\bm{R} = \bm{E} \bm{V}.
\label{R=EV}
\end{align}
\end{lemma}

As described in our previous work~\cite{YutakaPIMRC2025}, the core of the Doppler-first method is Lemma~1, which separates the joint estimation of Doppler and delay. Specifically, via the Prony method~\cite{weiss1963prony}, Doppler frequencies \( f_{D,p} \) are first estimated from the matrix \( \bm{E} \), which depends only on \( \{ f_{D,p} \} \). This is why the method is called \emph{Doppler-first}.
Once the Doppler component is extracted, the matrix $\bm{V}$ can be obtained. By removing the Doppler effect, one can obtain the matrix $\bm{\widetilde V_{p,\ell}}$, which contains only the delay parameter and is expressed as
\begin{align}
    \widetilde V_{p,\ell} = \alpha_p s(\ell T_s - t_{d,p}).
\end{align}
%
Here we use the fact that the transmitted signal (\ref{s(t)}) is accurately approximated by the Dirichlet waveform (\ref{dirichlet_waveform}).
The $U_tM$-point DFT of $\widetilde V_{p,\ell}$ ($\ell = 0,1,\ldots, U_tM-1$) is then given by 
\begin{equation}
\begin{aligned}
& \sum_{\ell =0}^{U_t M-1} \widetilde{V}_{p,\ell} \, e^{-j\frac{2\pi}{U_t M} m\ell} \\
&= 
\begin{cases}
\alpha_p U_t M  e^{- j2\pi m \frac{t_{d,p}}{T} }, \\
\quad\quad \quad \mbox{if } -\frac{M}{2} \le m \le \frac{M}{2}-1 \mbox{ mod } U_t M,\\
0, \quad \quad\mbox{otherwise}.  
\end{cases}
\label{DFT_of_V_tilde}
\end{aligned}
\end{equation}
From this, $t_{d,p}$ can also be estimated using the Prony method.

Based on the above analysis, a primal Prony-based two-stage method is proposed, in which the first stage performs Doppler estimation. The details are given below.

\subsubsection{Stage 1: Doppler Estimation}
Let $\bm{R}$ be the received signal matrix with $U_t M$ columns, and let $\bm{T}$ be the transpose of $\bm{R}$ with the columns reversed, i.e.,
\begin{align}
T_{\ell, n} = R_{N-1-n, \ell},
\end{align}
where $\quad \ell=0,\dots,U_t M-1,$ and $\ n=0,\dots,N-1$.
Prony's method is applied to each row of $\bm{T}$ to estimate the Doppler frequencies, 
and all resulting equations share the same solutions. 
This leads to $U_t M$ simultaneous equations for more robust estimation. 
The method determines a vector $\bm{a} = (a[0], a[1], \ldots, a[\hat{P}])^t$ with $a[0]=1$ that satisfies $\bm{T} \bm{a} = \bm{0}$, 
where $\hat{P} = N-1$ and $\bm{0}$ is a zero vector of dimension $U_t M$.

Based on obtained $\bm{a}$, the zeros of the polynomial,
\begin{align}
a[0] x^{\hat{P}} + a[1] x^{\hat{P}-1} + \cdots + a[\hat{P}-1] x + a[\hat{P}] = 0,
  \label{polynomial}
\end{align}
are denoted by $Z_p$, $p=1,\ldots,\hat{P}$, and the estimated Doppler frequencies are then extracted as
\begin{align}
\hat{f}_{D,p} = \frac{\arg(Z_p)}{2\pi T},
\end{align}
where $\arg(Z_p)$ denotes the phase angle of $Z_p$ in $[-\pi, \pi]$.

\subsubsection{Preprocessing before Stage 2}

The matrix $\bm{\hat{E}} = \left( e^{j2\pi \hat{f}_{D, p} n T} \right)_{n,p}$ is first reconstructed from the estimated Doppler shifts $\hat{f}_{D,p}$, and the matrix $\bm{V}$ is subsequently calculated as
\begin{align}
    \hat{\bm{V}} = \arg \min_{ \bm{V} } \left\| \bm{R} - \bm{\hat{E}} \bm{V} \right\|^2 
    = \bm{\hat{E}}^{\dagger} \bm{R},
\end{align}
where $\dagger$ denotes the Moore–Penrose pseudo-inverse.

Next, the Doppler effect in $\hat{\bm{V}}$ is eliminated by
\begin{align}
    \widetilde{V}_{p,\ell} = \hat{V}_{p,\ell} \, e^{-j2\pi \hat{f}_{D,p} \ell T_s}.
\end{align}

After that, a $U_t M$-point DFT is applied to $\widetilde{V}_{p,\ell}$ for each $p$:
\begin{align}
    Y_p[m] = \sum_{\ell=0}^{U_t M-1} \widetilde{V}_{p,\ell} \, e^{-j \frac{2\pi}{U_t M} m \ell}.
\end{align}
If $\hat{f}_{D,p}$ are accurate and noise-free, (\ref{DFT_of_V_tilde}) gives
\begin{align}
    Y_p[m] = \alpha_p U_t M \, e^{-j2\pi m t_{d,p} / T},
    \label{Y_p[m]}
\end{align}
for $m \in \left\{-\frac{M}{2}, \dots, \frac{M}{2}-1 \right\} \ \mathrm{mod}\ U_t M$.
\subsubsection{Stage 2: Delay Estimation}
Since the path is expected to be sparse, we assume $L=1$ in practice, where $L$ is the number of paths corresponding to the same Doppler shift. 
For each vector $\bm{Y}_p = (Y_p[m])_{m=-M/2}^{M/2-1}$, we apply the Prony method again by constructing a Toeplitz matrix
\begin{align}
    (\bm{T}_p)_{ij} = Y_p\left[L - M/2 + i - j\right],
\end{align}
where $i = 1, \ldots, M - 1$ and $j = 1, 2$.

Similarly, we solve the nonzero vector $\bm{a} = (a[0], a[1])^\mathrm{T}$ satisfying $\bm{T}_p \bm{a} = \bm{0}$, and compute the roots, denoting them by $Z_p$, of the polynomial in (\ref{polynomial}) with $\hat{P}$ replaced by $L=1$. 

Finally, the estimated time delays corresponding to the Doppler shift $\hat{f}_{D,p}$ are given by
\begin{align}
    \hat{t}_{d,p} = T \left( -\frac{\arg Z_p}{2\pi} \bmod 1 \right),
    \label{hat_t}
\end{align}
where the modulo operation ensures $\hat{t}_{d,p} \in [0, T)$.

\subsection{Primal Two-Stage Prony Method: Delay-First}\label{sect:dual}
The Doppler-first method was introduced in the previous section. Due to symmetry, the delay-first method can also perform the estimation, with its core outlined in Lemma~2.
\begin{lemma}
Let $\bm{R}' = ( R'_{m,k} )$
be an $M\times (U_f N)$ matrix defined by
\begin{align}
R'_{m,k} = R_{\rm FD}[ m U_fN+k],
\end{align}
where $m \in \left[ -\frac{M}{2}, \frac{M}{2}-1 \right]$ and $k \in \left[ -\frac{U_f N}{2}, \frac{U_f N}{2}-1 \right]$.

Subsequently, we respectively define the $M \times P$ matrix $\bm{E}' = (E'_{m, p})$ and $P \times (U_f N)$ matrix $\bm{V}' = (V_{p,k}')$ by
\begin{align}
E'_{m, p} = e^{ -j2\pi t_{d,p} m/T},
\end{align}
\begin{align}
V_{p,k}' = \alpha_p S\left( \frac{k}{U_f NT} - f_{D,p} \right) e^{ - j2\pi t_{d,p} \frac{k}{U_f NT}}.
\end{align}

Then, we have a result similar to Lemma~1, given by
\begin{align}
\bm{R}' = \bm{E}' \bm{V}'.
\label{R'=E'V'}
\end{align}
\end{lemma}

Similar to the Doppler-first method, $\bm{T}'$ is constructed as 
\begin{align}
T'_{k,m} = R'_{M-1-m, k},
\end{align}
where $m = 0, 1, \ldots, M-1$ and $k = 0, 1, \ldots, U_fN-1$.
We then find the vector $\bm{a}$ satisfying $\bm{T}' \bm{a} = \bm{0}$.
With $\hat P = M-1$, the zeros, denoted by $Z_p$, of the polynomial (\ref{polynomial}) associated with $\bm{a}$ yield the delays $\hat t_{d,p}$ via (\ref{hat_t}).
Next, we compute
\begin{align}
\hat{\bm{E}}' &= \big( e^{-j 2\pi \hat t_{d,p} m/T} \big)_{m,p}, \\
\hat{\bm{V}}' &= ( \hat{\bm{E}}' )^\dagger \bm{R}', \\
\widetilde{V}_{p,k}' &= \hat V_{p,k}' \cdot e^{j2\pi \hat t_{d,p} k/(NT)}, \\
Y_p'[n] &= \sum_{k=0}^{U_fN-1} \widetilde{V}_{p,k}' e^{j \frac{2\pi}{U_fN} nk}, 
\end{align}
and finally obtain $\hat f_{D,p}$ corresponding to $\hat t_{d,p}$, following the same procedure as in Section~\ref{sect:primal}.
\subsection{Parallel method: integration for two estimates}
The Doppler-first method may fail to separate paths with similar Doppler but different delays. Conversely, the delay-first method may also struggle when delays are close but Dopplers differ. Integrating the path candidates from both methods mitigates these limitations.

Let $(\hat t_{d,p}^{(1)}, \hat f_{D,p}^{(1)})$, $p=0,1,\ldots, N-1$, be the estimation of the Doppler-first method, and $(\hat t_{d,p}^{(2)}, \hat f_{D,p}^{(2)})$, $p=0,1,\ldots, M-1$ be the estimation of the delay-first. Let $\delta_t, \delta_f \in [0, 0.5)$ be the time-domain and frequency-domain resolutions, which are proportional to the noise level. If the pair $(p,i) \neq (q,j)$ meets 
$\bigl| \hat t_{d,p}^{(i)} - \hat t_{d,q}^{(j)} \bigr| < \delta_t T \quad \text{and} \quad 
\bigl| \hat f_{D,p}^{(i)} - \hat f_{D,q}^{(j)} \bigr| < \delta_f / T$,
the original estimates will be replaced by $\frac{\hat t_{d,p}^{(i)} + \hat t_{d,q}^{(j)}}{2} \quad \text{and} \quad 
\frac{\hat f_{D,p}^{(i)} + \hat f_{D,q}^{(j)}}{2}$. This merging continues until no such pair remains.

Next, let $\Theta$ denote the remaining delay-Doppler pairs with cardinality $\tilde P = |\Theta|$. Define
\begin{align}
\hat r_p[\ell] = s(\ell T_s - \hat t_{d,p} ) e^{j 2\pi \hat f_{D,p} \ell T_s},
\end{align}
where $\ell = -N_0 U_t M, \dots, (N+1+N_0) U_t M - 1$. The received vector is
$\bm{r}_{\rm TD} = \sum_{p=1}^P \alpha_p \bm{r}_p + \bm{z}$,
where $\bm{z}$ is the noise vector. 
In this condition, the $\bm{\alpha}$ can be estimated via the least-squares, 
\begin{align}
\hat {\bm{\alpha}} = \arg\min_{\bm{\alpha}} 
\lVert \bm{r}_{TD} - \sum_{p=1}^{\tilde P} \alpha_p \hat{\bm{r}}_p \rVert^2.
\end{align}
Paths with small power, $\lvert \alpha_p \rvert < \delta_\alpha \times \max_{p'} \lvert \alpha_{p'} \rvert$, are discarded from $\Theta$. The threshold $\delta_\alpha \in (0, 1)$ depends on the noise level.

\section{Simulation results}

\begin{figure}
    \centering
    \includegraphics[width=0.8\linewidth]{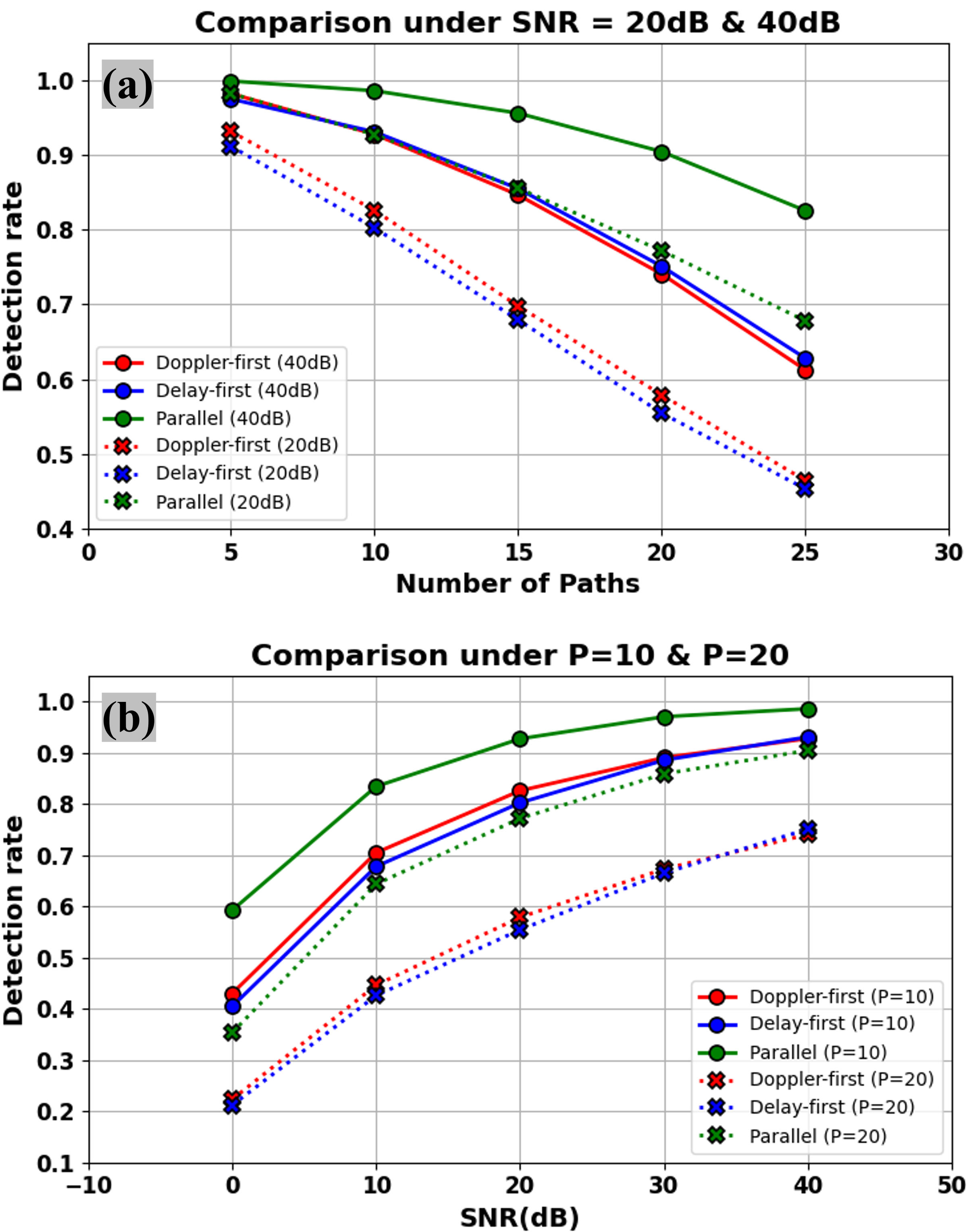}
    \caption{Comparison of the delay-first, Doppler-first, and parallel methods under the same random seed. 
}
    \label{compare}
\end{figure}

The simulations were performed with 
\(N=M=32\), using a fixed random seed. $\delta_\alpha$, $\delta_t$, and $\delta_f$ are set to 0.01, 0.1, and 0.1, respectively. 
The delay and Doppler are uniformly distributed over $[0,T]$ and $[-T/2,T/2]$.
The number of Monte Carlo iterations is $\mathrm{Runs}=1000$. A path is considered successfully matched if the delay and Doppler estimation deviations are less than $0.5T$ and $0.5/T$. The detection rate is $R_{\mathrm{detection}}=\frac{D}{\mathrm{Runs}\times P}$ where $D$ is the number of detections and $P$ is the number of paths.
The results, summarized in Fig.~\ref{compare}, highlight the outperformance of the parallel method:


\begin{itemize}
\item \textbf{Scalability with Path Number:}
Fig.~\ref{compare} (a) demonstrates that the parallel method is relatively unaffected by the increase in path numbers and consistently achieves higher detection rates than the other two methods. Its stable performance under both $20$ and $40$ dB illustrates its superior scalability, particularly in dense multipath scenarios.

\item \textbf{Low-SNR Robustness:} 
As shown in Fig.~\ref{compare} (b), the parallel method exhibits greater robustness to noise compared with the other two approaches. Even at low SNR, it maintains higher detection rates across different path numbers, confirming its reliability under challenging conditions.

\end{itemize}



    
    
However, the parallel method may still face challenges under strong noise and dense multipath conditions, suggesting potential for further refinement. 
Nevertheless, it provides more robust and accurate estimates, making it a promising solution for high-precision applications such as V2V and ISAC.

\section{Concluding Remarks}
We proposed a Prony-based parallel two-stage method for estimating the delays and Doppler shifts of doubly selective multipath channels using OTFS pilot signals. The method combines two sub-approaches: one estimating Doppler-first and the other estimating delay-first. Numerical simulations demonstrated that the proposed method outperforms each approach, confirming its effectiveness in accurately estimating multipath channel parameters.

\bibliographystyle{IEEEtran}
\bibliography{mybibliography.bib, gyoseki.bib, complement.bib}

\appendices

\end{document}